\pdfoutput=1
\documentclass{elsart}
\usepackage{graphicx,color}
\usepackage{amsmath,amssymb}
\usepackage{slashbox}
\usepackage{epstopdf}

\def\cmp#1#2#3{Comm. Math. Phys. #1 (#3) #2}

\def\ibid#1#2#3{{\it ibid.} #1 (#3) #2}

\def\npa#1#2#3{Nucl. Phys. A #1 (#3) #2}
\def\npb#1#2#3{Nucl. Phys. B #1 (#3) #2}

\def\plb#1#2#3{Phys. Lett. B #1 (#3) #2}
\def\prc#1#2#3{Phys. Rev. C #1 (#3) #2}
\def\prd#1#2#3{Phys. Rev. D #1 (#3) #2}
\def\prl#1#2#3{Phys. Rev. Lett. #1 (#3) #2}

\def\ptp#1#2#3{Prog. Theor. Phys. #1 (#3) #2}
\def\ptps#1#2#3{Prog. Theor. Phys. Supp. #1 (#3) #2}

\def\rmtp#1#2#3{Rev. Math. Phys. #1 (#3) #2}

\newcommand{\Slash}[1]{{\ooalign{\hfil/\hfil\crcr$#1$}}}
\newcommand{\tr}{{\rm tr}}
\newcommand{\Nc}{N_{\rm c}}
\newcommand{\Nf}{N_{\rm f}}

\newcommand{\lqcd}{\Lambda_{\rm QCD}}

\newcommand{\ra}{\rangle}
\newcommand{\la}{\langle}
\newcommand{\up}{\! \uparrow }
\newcommand{\down}{\! \downarrow }
\newcommand{\fpi}{f_\pi}

\begin{document}
\begin{frontmatter}
\title{The Dichotomous Nucleon: 
Some Radical Conjectures for the Large $\Nc$ Limit}
\author[kyoto]{Yoshimasa Hidaka},
\author[riken]{Toru Kojo},
\author[riken,bnl]{Larry McLerran}, and
\author[bnl]{Robert D. Pisarski}

\address[kyoto]{Department of Physics, Kyoto University, Sakyo-ku, Kyoto
606-8502, Japan}
\address[riken]{RIKEN/BNL Research Center, Brookhaven National
  Laboratory,\\ Upton, NY-11973, USA}
\address[bnl]{Department of Physics, Brookhaven National Laboratory, Upton,
  NY-11973, USA}

\maketitle

\begin{abstract}
We discuss some problems with
the large $\Nc$ approximation for nucleons which arise
if the axial coupling of the nucleon to pions is large, $g_A \sim \Nc$.
While $g_A \sim \Nc$ in non-relativistic quark and Skyrme models,
it has been suggested that Skyrmions may collapse to a small size,
$r \sim 1/\fpi \sim \lqcd^{-1}/\sqrt{\Nc}$.
(This is also the typical scale over which the string vertex moves
in a string vertex model of the baryon.)
We concentrate on the case of two flavors, where we suggest
that to construct a nucleon with a small axial coupling,
that most quarks are bound into 
colored diquark pairs, which have zero spin and isospin.
For odd $\Nc$, this leaves one unpaired quark,
which carries the spin and isospin of the nucleon.
If the unpaired quark is in a spatial wavefunction orthogonal to the
wavefunctions of the scalar diquarks, then
up to logarithms of $\Nc$, the unpaired quark only costs
an energy $\sim \lqcd$.  This naturally
gives $g_A \sim 1$ and has other attractive features.
In nature, the wavefunctions of the paired and unpaired quarks 
might only be approximately
orthogonal; then $g_A$ depends weakly upon $\Nc$.
This dichotomy in wave functions could arise 
if the unpaired quark
orbits at a size which is parametrically large in comparison to that
of the diquarks.  
We discuss possible tests of these ideas from numerical simulations on the 
lattice, for two flavors and
three and five colors; the extension of our ideas to more than
three or more flavors is not obvious, though.
\end{abstract}


\end{frontmatter}

\section{Introduction}

The large $\Nc$ limit of 't Hooft \cite{'tHooft:1973jz} for the description of
baryons has been developed by Adkins, Nappi and Witten \cite{Adkins:1983ya}.  
In this limit, the nucleon is a topological excitation of the pion field, 
where the
pion field is described by a non-linear sigma model plus a Skyrme
term \cite{Skyrmion}. This topological excitation 
is described by a stable
soliton solution of size $r \sim 1/\lqcd$, which
is a Skyrmion; $\lqcd$ is a mass scale typical of the strong
interactions.

The action of the Skyrmion is $\sim \Nc$,
and so it contains of order $\Nc$ coherent pions.  
In the Skyrme model, the nucleon pion coupling constant is 
enhanced from its naive value, 
$g_{\pi NN} \sim \sqrt{\Nc}$, which arises from
counting the number of quarks inside a nucleon, 
to become $g_{\pi NN} \sim \Nc^{3/2}$. 
This is a consequence of the coherent nature of the pions which
compose the Skyrmion. 
By the Goldberger-Treiman relation~\cite{Goldberger:1958vp}, 
the axial coupling $g_A$ is then of order $\Nc$.
Such a strong axial coupling 
generates strong spin-isospin dependent
forces, of order $\Nc$, 
out to distances which are large in comparison 
to the size of the nucleon, $\sim 1/\lqcd$.  
In the limit of massless pions, 
these interactions are of infinite range.
In Monte-Carlo computations of the nucleon-nucleon force on the lattice, 
no strong long range tails
are seen; indeed, even at intermediate ranges the forces do not appear to be
large \cite{Ishii:2009zr} 
(Some cautions on the interpretation of lattice results
were raised in \cite{Beane:2009py}).
In addition, the magnetic moment of
the proton would be of order $\Nc$, 
which would also 
generate strong electromagnetic interactions \cite{Karl:1984cz}.

Such a description of the nucleon at infinite $\Nc$
appears to be rather different from what we observe for $\Nc = 3$.
At finite $\Nc$, these problems might be fixed by a fine
tuning of parameters.   For example, in the Skyrme model description of Ref.
\cite{Adkins:1983ya}, the parameter $1/e^2$ that controls the strength of the
Skyrme term, and which stabilizes the Skyrmion at a non-zero radius, 
should be of order $\Nc$.
To provide a  phenomenologically viable description of the nucleon
for $\Nc =3 $, though, it is taken to be $3.3 \times 10^{-2}$.

Another generic problem is the nature of nuclear matter.  Some of the channels
for the long distance spin-isospin dependent forces are attractive.  This means
that the ground state of nuclear matter is a crystal and the binding energy is
of order $\Nc \lqcd$ \cite{skyrme_crystal}.
On the other hand, ordinary nuclear matter is very weakly bound, with a binding
energy $\delta E  \sim 16$~MeV \cite{Brown_book}.  
This number seems to be closer to $\lqcd/\Nc$
than to $\Nc \lqcd$, the value typical of a Skyrme crystal. 
Moreover, nuclear matter appears to be in a liquid state, and
not a crystal.

An excellent discussion of the properties of the nucleon-nucleon force is found
in Ref. \cite{Witten:1979kh,Dashen:1993as,Kaplan}.  Many of the relationships
derived there are generic relationships between the magnitudes of various
forces, and these seem to work quite well.
Thus it is somewhat of a mystery why
the large $\Nc$ limit for baryons can work well in some contexts, but provide
qualitative disagreement in others.

Yet another problem is the mass splitting between the
nucleon and $\Delta$.
Consistency conditions at large $\Nc$
and standard large $\Nc$ counting indicate that this mass difference
is $\sim \lqcd/\Nc$ \cite{Witten:1979kh,Dashen:1993as}.  
In QCD, though, it is $\sim 300$~MeV, 
which is $\sim \lqcd$.

A large value of $g_A$ also generates problems in writing
a chiral effective theory for the nucleon.  In the linear sigma model, chiral
symmetry implies that there is a large coupling to the sigma
meson,  $g_{\sigma NN} = g_{\pi NN} \sim \Nc^{3/2}$.  
Such a large coupling generates
self-energy corrections to the nucleon that would be larger
than $\Nc$.  In addition, if the axial coupling is of
order $\Nc$, self-interactions associated with an axial-vector
current should result in a 
significant contribution to the nucleon mass.  
If there is some way to lower
the axial coupling, which does not greatly increase the
mass of the nucleon,
then it is plausible that nature would realize this possibility.

Ultimately, large self-energies for the nucleon might destabilize
a nucleon of size
$\sim 1/\lqcd$.  One might be tempted to argue that this
cannot happen in QCD, since the action in QCD is of order $\Nc$, 
and a collapsed
soliton, with a size other than $\lqcd$, should
have a mass which is not linear in $\Nc$.  This would be a
strong argument if the nucleon appeared as a purely
classical solution of the QCD equations of motion, as a Skyrmionic soliton for
example.  
Following others, however, we suggest that the Skyrmionic soliton may
collapse 
\cite{mackenzie,Aitchison:1986aq,Aitchison:1985yv,Aitchison:1984ys,Ripka:1985am}.
If so, at short distances the nucleons are 
more naturally described by quarks
rather than by coherent pions.
The quarks cannot collapse to a small size without
paying a price of order $\Nc/R$ in quark kinetic energy.   
The relevance of quark descriptions inside of the nucleon
was also emphasized in \cite{Diakonov:1987ty}.

A key observation in this paper 
is that such constituent quarks are the main origin of the
axial charge $g_A$, which is the source of pion fields.
If $\Nc$ constituent quarks totally have a small axial charge, 
$g_A \sim 1$,
then the problems related to large coherent pions
will be solved.

We suggest such a nucleon wavefunction.
Most quarks are bound into colored diquarks \cite{jaffe}. 
For odd $\Nc$, that leaves one unpaired quark.  
We then put that unpaired quark into a wavefunction which is 
approximately {\it orthogonal} to those of the paired quarks.
This can be
accomplished by making the spatial extent of the unpaired quark 
larger than that of the paired diquarks: it is ``dichotomous''.
Putting the
additional quark into such a wavefunction costs an energy of order
$\lqcd$, up to logarithms of $\Nc$ (as we show later). 
Such a construction results in small self-energies 
from the pion-nucleon self-interactions, as a result of  
$g_A \sim 1$.
It is also clear that long-range
nucleon-nucleon interactions are no longer strong.

This is a minimal modification of the
naive non-relativistic quark model of the nucleon.
There quarks are paired
into diquarks, save for one quark that carries the quantum numbers of the
nucleon.  It is usually assumed, however, that 
{\it all} of the quarks, paired or not, have the same spatial wavefunction.  
This gives $g_A =(\Nc+2)/3$, and the problems discussed 
above \cite{Karl:1984cz,Kakuto:1981ei}.

A trace of the collapsed Skyrmion might appear at a scale size of order
$1/\fpi$.  This size corresponds to the intrinsic scale of a quantum
pion.  Since $\fpi \sim \sqrt{\Nc} \lqcd$ at large $\Nc$,
the size of the nucleon shrinks to zero as $\Nc \rightarrow \infty$.
We will also show that such a small size naturally 
arises in a string vertex model of the
nucleon, as the root mean square 
fluctuations in the position of the string vertex.
Of course, the contribution of the string vertex to the mass of the nucleon is
of order $\fpi$, as most of the mass of the nucleon is generated by a cloud of
quarks and quark-antiquark pairs surrounding the collapsed Skyrmion, or string
vertex.  The picture we develop has some aspects in common with
bag models \cite{Chodos:1974je}, and particularly the hybrid descriptions of
Brown and Rho \cite{Brown:1979ui,Thomas:1981vc}.  

The collapsed Skyrmion we conjecture has 
some features which are similar to the 
nucleon in the Sakai-Sugimoto model \cite{Sakai:2004cn}. 
They suggest that the Skyrmion, computed in the action to 
leading order in strong coupling, is
unstable with respect to collapse.  It is stabilized by $\omega$ vector meson
interaction, which is of higher order correction in strong coupling.  
It is argued
that the nucleon has a size of order $1/(\sqrt{g^2 \Nc} \lqcd)$.  The methods
used to derive this result are questionable at sizes
$\ll 1/\lqcd$, but at least this shows that there is a small
object in such theories.  It is quite difficult for  $\omega$ exchange 
or other strong
coupling effects to stabilize the nucleon once it acquires a 
size much less than
$1/\lqcd$.  The basic problem is that mesons will decouple from small
objects due to form factor effects.  
Without form factors, the $\omega$ interaction
generates a term $\sim 1/R$, which resists collapse;
form factors convert this into a factor of $\sim R$, which is harmless as $R$
shrinks to zero.

The outline of this paper is as follows:  In Sec. 2 we review the
sigma model and its predictions
for nucleon structure.  We show that its predictions for the large $\Nc$
properties of the nucleon are at variance from the large $\Nc$ limit predicted
for a Skyrmion of size $\sim 1/\lqcd$.   In Sec. 3 we discuss
the general form of nucleon-nucleon interactions in the sigma model and in the
Skyrme model.  In Sec. 4 we argue that the Skrymion might
collapse to a size scale of order $1/\fpi$
\cite{mackenzie,Aitchison:1986aq,Aitchison:1985yv,Aitchison:1984ys,Ripka:1985am}.  
In Sec. 5 we discuss
the string vertex model of Veneziano \cite{Rossi:1977cy}. 
In particular, we
argue that the spatial extent of the string vertex is typically of order
$1/\fpi$, which is the minimal size for the string vertex.  
Such a vertex might
be thought of as the localization of baryon number.
Quarks attached to the ends of strings will nevertheless have a spatial extent
of order $1/\lqcd$ to avoid paying a huge price in quark kinetic
energies.  In Sec. 6 we compute the contribution to $g_A$ arising
from quarks.  Using the non-relativistic quark model, we find that if we make
the wavefunction of those quarks paired as diquarks,
and that of the unpaired quark
have a small overlap, then $g_A$ is
parametrically smaller in $\Nc$
than the canonical value of $g_A = (\Nc+2)/3$.  
An explicit computation of $g_A$
and the magnetic moments for such a variable overlap
is carried out in Appendix A and B.   
In Sec. 7 we present arguments about
how, dynamically, such a small overlap might be achieved.
In Sec. 8 we summarize our arguments, and discuss how they
might be tested through numerical simulations on the lattice.

\section{The Sigma Model}

Let us begin by reviewing how the long range 
nucleon-nucleon interaction depend
on $\Nc$ in the  sigma model.  The linear sigma model is written in the form
\begin{eqnarray}
 S & = &  \int~d^4x~ \left\{{1 \over 2} \left( \partial_\mu \sigma \partial^\mu
\sigma + \partial_\mu \pi^a \partial^\mu \pi^a \right)  
-{\mu^2 \over 2} (\sigma^2 +
(\pi^a)^2) + {\lambda \over 4} (\sigma^2 + (\pi^a)^2)^2  \right.\nonumber \\
 & & \left.  +{\overline \psi} 
(-i \Slash{\partial} + g(\sigma +i \pi \cdot \tau \gamma^5 ) )\psi
\right\} ,
\end{eqnarray}
where $\psi$ denotes the nucleon field.
Our metric convention is $g^{00} = -1$.
The naive arguments of large $\Nc$ QCD would have the mass term $\mu$ of order
$\lqcd$,
the four meson coupling $\lambda \sim {1 / \Nc}$, and the pion nucleon
coupling $g \sim \sqrt{\Nc}$.

Upon extremizing the action,  we find that $M_\sigma \sim \mu$, $M_\pi = 0$,
$M_{N} \sim g \mu/\sqrt{\lambda} \sim \Nc \mu$.  Therefore the typical
large $\Nc$ assignments of couplings are consistent with the 
nucleon mass being of
order $\Nc$, the sigma mass of order one, and a weakly coupled pionic and sigma
system.  Note that the sigma is strongly coupled to the nucleon, 
consistent with large $\Nc$ phenomenology.

What about the pion coupling?  It is naively of order $\sqrt{\Nc}$ but the
$\gamma^5$ matrix, because of the negative parity of the pion, 
suppresses pion emission
when the momentum of the pion is much less than that of the nucleon.  A non
relativistic reduction of the pion nucleon interaction gives
\begin{equation}
      g \pi_a {\overline \psi} \tau^a \gamma^5 \psi \sim {g \over {2 M_N}}
(\partial_\mu \pi^a) {\overline \psi}
      \tau^a \gamma^\mu \gamma^5 \psi .
\end{equation} 
This equation means that one pion emission is not of order $\sqrt{\Nc}$ at long
distances, but of order $1/\sqrt{\Nc}$.
Thus the potential due to one pion exchange is of order $1/\Nc$,
and not of order $\Nc$.

\begin{figure}[htbp]
\begin{center}
\includegraphics[width=0.40\textwidth] {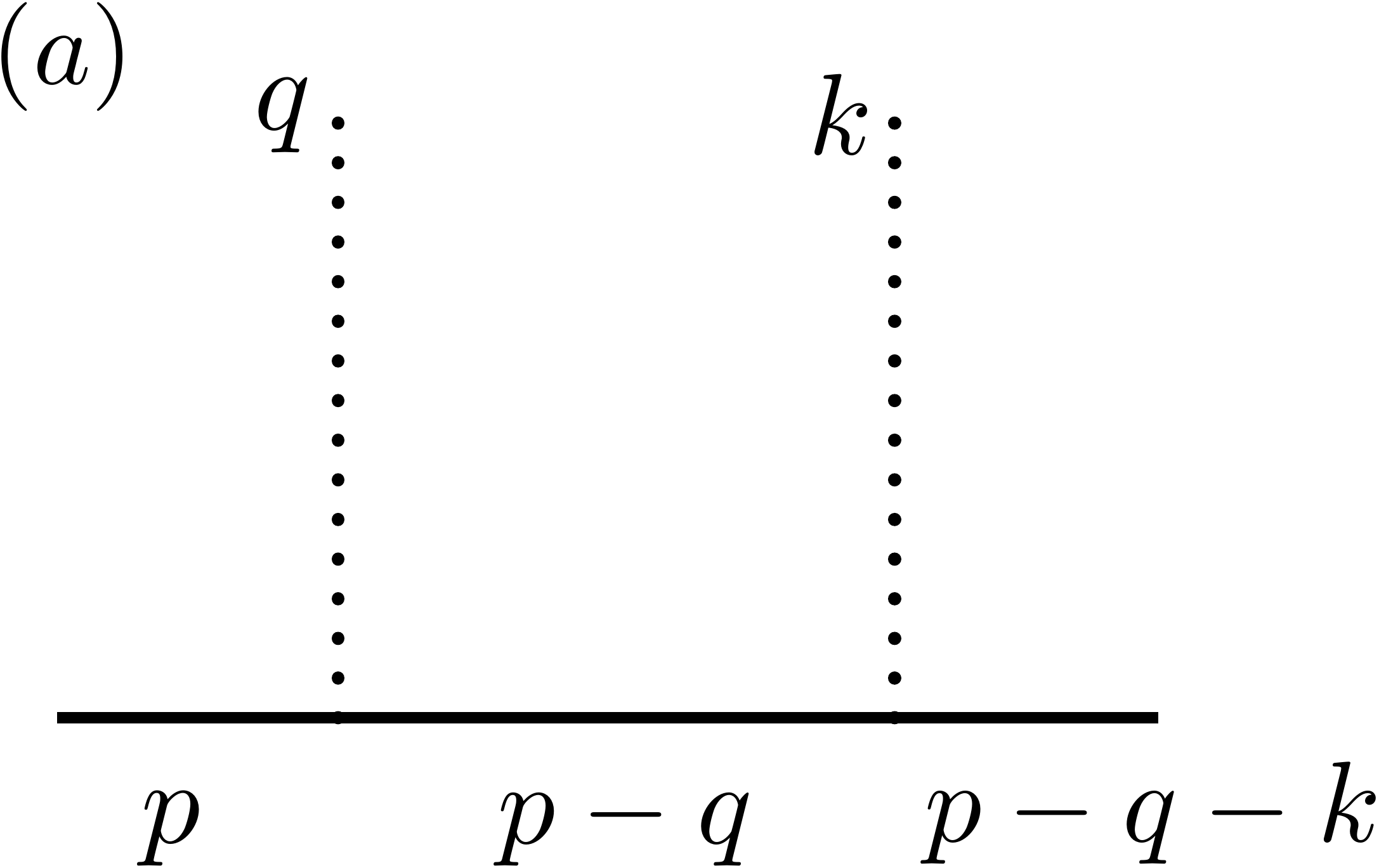}  
\qquad
\includegraphics[width=0.40\textwidth] {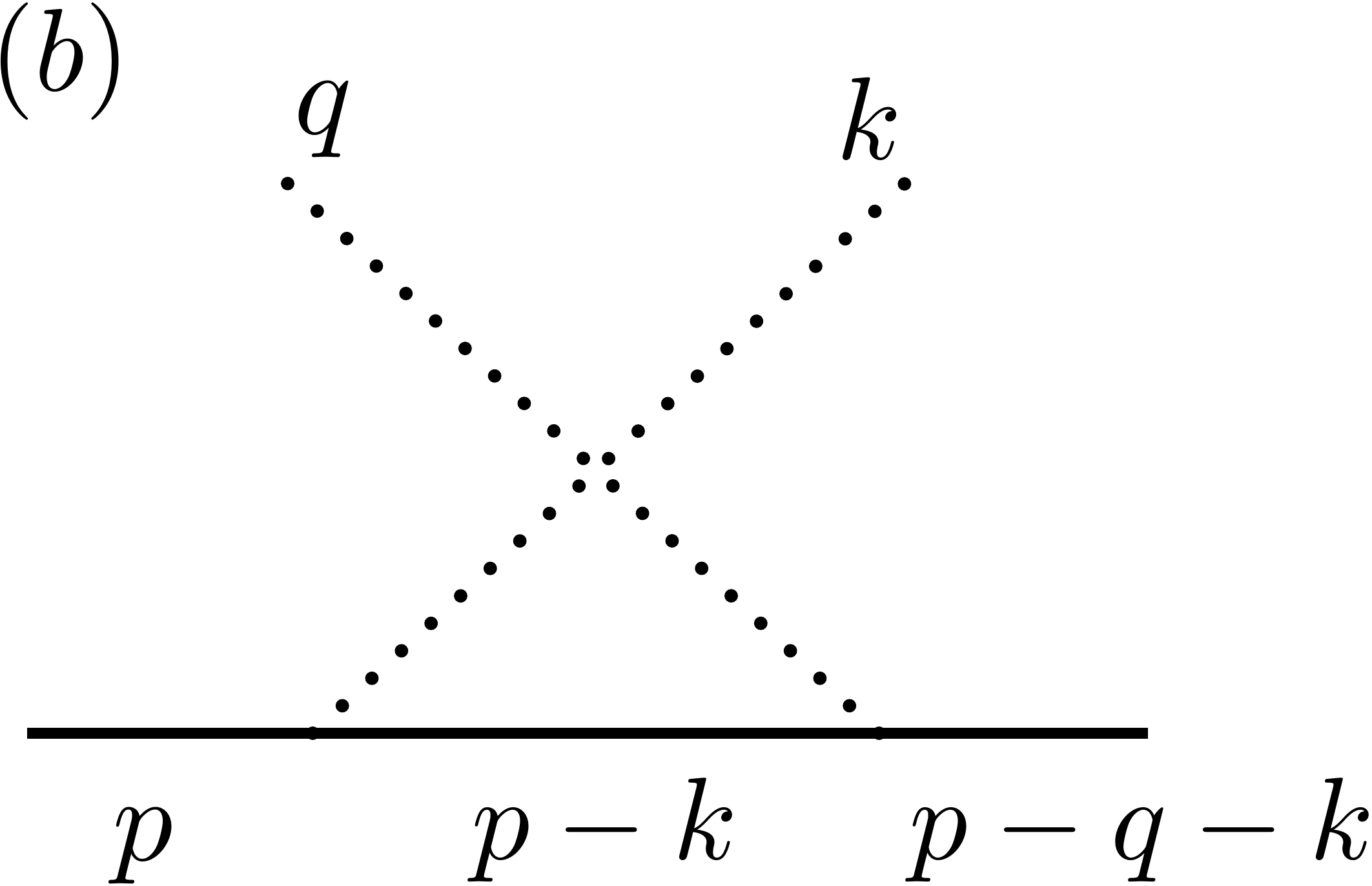}
\end{center}
\caption{a:  One of the two pion exchanges.  b:  The crossed diagram.}
\label{pions}
\end{figure}

\begin{figure}[htbp]
\begin{center}
\includegraphics[width=0.40\textwidth] {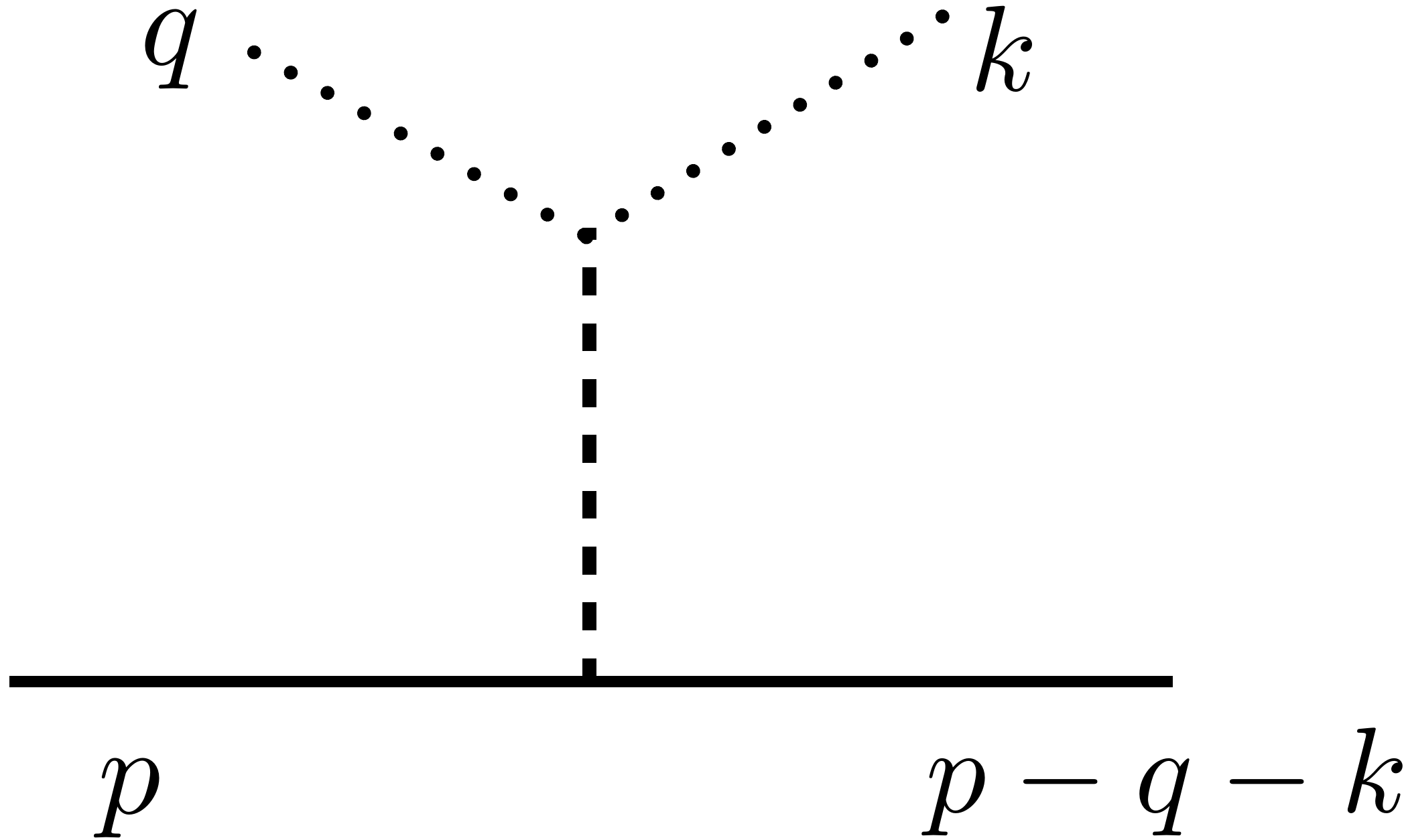}  
\end{center}
\caption{ Two pions produced by sigma exchange.}
\label{pionstoo}
\end{figure}

One might object that in higher orders 
this is not true, since one might expect 
the non-relativistic decoupling of the pions would disappear when one considers
two pion exchange.  If one considers the diagram
in Fig. \ref{pions}, this contribution is naively of order $\Nc$.  
The sum of the two diagrams cancel to leading order when 
$ q, k \ll M$, making it
again naively of order one.  However, when the diagram in 
Fig. \ref{pionstoo} is
included, which is also of order one in powers of $\Nc$, 
there is a cancellation
with the above two diagrams when the momentum of the pions is small compared to
$\mu$.  When all is said and done, we conclude that for momentum small compared
to the QCD scale, the interaction is of order $1/\Nc$.  This corresponds to a
suppression of $1/\sqrt{\Nc}$ for each pion emitted.

In fact, Weinberg proves by an operator transformation on the sigma model
action, that this cancellation
persists to all orders in perturbation in the theory, and that pion emission
when soft is always suppressed by $1/\sqrt{\Nc}$ for each emitted
pion \cite{weinberg}.

This conclusion about the strength of  the nucleon force is 
consistent with what
we know about nuclear matter.  Nuclear matter is weakly bound, 
and has a binding
energy which is of order $\Lambda^2/M_N \sim 1/\Nc$.  Such a parametric
dependence on $\Nc$ is seen in nuclear matter computations where pion exchange
is augmented by a hard core interaction \cite{weise}.  
The hard core interaction presumably arises when
momentum transferred is of order $\lqcd$, and interactions become of
order one in powers of $\Nc$.  In nuclear matter computations, the hard core
essentially tells the nucleons they cannot go there, and 
its precise form is not
too important.

It is useful to consider the non-linear sigma model, as this is the 
basis of the
Skyrme model treatment.  The non-linear sigma model is essentially the infinite
sigma particle mass limit of the
linear sigma model.  It should be valid at distance scales much larger than
$1/\lqcd$,
which is also the range of validity of the linear sigma model.  The action for
the non-linear sigma model is
\begin{equation}
  S = \int~d^4x~\left\{{\fpi^2 }~ 
\tr ~\partial_\mu U \partial^\mu U^\dagger  +
  \overline \Psi \left( -i \Slash{\partial} + M \overline U
\right) \Psi \right\} .
\end{equation}
In this equation,
\begin{equation}
  U = e^{i {\tau \cdot \pi }/\fpi} ,
\end{equation}
and
\begin{equation}
  \overline U = e^{i {\tau \cdot \pi \gamma^5 }/\fpi} ,
\end{equation}
where $\fpi \sim \sqrt{\Nc} \lqcd$, and the nucleon mass $M \sim \Nc
\lqcd$.

Weinberg's trick is to rotate away the interactions in the mass 
term by a chiral rotation,
\begin{equation}
  \overline U \rightarrow V^{-1/2} \overline U V^{-1/2} = 1 .
\end{equation}

After this rotation of the nucleon fields, the action becomes,
\begin{equation}
 S = \int~d^4x~\left\{{\fpi^2 }~\tr~ \partial_\mu U \partial^\mu U^\dagger  +
  \overline \Psi \left( {1 \over i} \gamma^\mu (\partial_\mu + \gamma^5 V^{1/2}
\partial_\mu V^{-1/2}) + M \right) \Psi \right\} .
\end{equation}
We do not need to know the explicit form of V to extract the essential 
physics. 
The point is that
the expansion in powers of the pion-nucleon interaction involves a factor of
$1/\sqrt{\Nc}$ for each power of the pion field.  This arises because the
coupling to the pions is a derivative coupling,
and to get the dimensions right each power of the derivative times the pion
fields must be accompanied by a factor of $1/\fpi$.  
Notice also that the first
term in the expansion of the pion field is $1/\fpi$, and couples to 
the nucleonic axial-vector current.  The nucleonic axial-vector current is one for free
fermions, and the interactions in this theory, corresponding to decreasing
powers
of $1/\sqrt{\Nc}$, do not change the parametric dependence upon $\Nc$.  

While in these models above $g_A \sim 1$, it is possible to obtain
$g_A \sim \Nc$ by the addition of further terms to the effective Lagrangian.
In the linear sigma model, consider adding a term
\cite{kaplan2,weinberg_book}
\begin{equation}
\frac{\widetilde{g}}{\lqcd^2} \left( 
\overline{\psi}_L \left( \Phi^\dagger  \slash \!\!\! \partial \Phi 
\right) \psi_L
\; + \;
\overline{\psi}_R\left( \Phi \slash \!\!\! \partial \Phi^\dagger 
\right) \psi_R 
\right) \; .
\label{non_ren}
\end{equation}
Here $\psi_{L,R}$ are chiral projections of the nucleon field,
and $\Phi$ transforms under $SU_L(2) \times SU_R(2)$.  This term
is non-renormalizable, with the coupling having an overall dimension
of inverse mass squared.  We take this mass scale to be $\lqcd$, so the
coupling $\widetilde{g}$ is dimensionless.  Taking $\Phi \sim f_\pi U$,
this term generates an axial vector coupling of the pion to the nucleon,
and $g_A \sim \Nc \widetilde{g}$; see, {\it e.g.},
Eqs. (19.5.48), (19.5.49), and (19.5.50) of Ref. \cite{weinberg_book}.

Thus if we allow the addition of non-renormalizable terms to the
linear sigma model, $g_A$ can be treated as a free parameter.  Our
point, however, is that if one takes $g_A \sim \Nc$, then 
at large distances, where the nucleon-nucleon interaction is 
determined by pion exchange, that the corresponding interactions are
strong, $\sim \Nc$.  While certainly logically possible, 
this does not agree with the phenomenology of nucleon scattering,
which sees no long range tails which are large in magnitude
\cite{Ishii:2009zr}.

\section{The General Structure of the Nucleon-Nucleon Force in the Sigma Model
and the Skyrme Model}

It is useful to compute the general form of the nucleon-nucleon force in both
the Skyrme model and the 
sigma model.  Let us first consider the Skyrme model,
\begin{equation}
  S_{\rm skyrme} = \int ~ \d^4x~ \left\{ {{\fpi^2} \over 16} ~\tr~ \partial_\mu U
\partial^\mu U^\dagger + {1 \over {32e^2}} 
~\tr~  [\partial_\mu U, \partial_\nu U]^2 \right\} .
\end{equation}
The last term in this equation is the Skyrme term.  It has a 
coefficient $1/e^2$
that is assumed to be of order $\Nc$, and is positive.  There is a topological
winding number in the theory, and this winding
number can be related to the total baryon number by an anomaly.  The nucleon
corresponds to
the solution with winding number one.  The size of the baryon is found to be
$R_{\rm baryon} \sim 1/\sqrt{e\fpi} \sim 1/ \lqcd$, and is independent of
$\Nc$.  If the Skyrme term were zero or negative, the solution would 
collapse to
zero size.  

The two nuclear force is derived by considering a two Skyrmion solution and
computing the energy
of separation~\cite{nuclearForce}. Since if we simply redefine scale sizes in the Skyrme action by
defining a dimensionless
pion field as $\pi^\prime = \pi/\fpi$, and rescaling coordinates by
$\lqcd$, the Skyrme
action becomes explicitly proportional to $\Nc$ when all dimensional quantities
are so expressed.
Therefore, the potential between to nucleons is of the form
\begin{equation}
  V_{\rm skyrme}(r) \sim {\Nc \over r} F_{\rm skyrme}(\lqcd r) .
\end{equation}
This is clearly inconsistent with the result one gets from the 
Weinberg action. 
Here the lowest order diagram which contributes at distances much larger than
$1/\lqcd$ is due to one pion emission.  Its strength is of order $1/r 
(r\fpi)^2 \sim 1/\Nc$.  In the Skyrme model this difficulty is evaded
by arguing that the strength 
of the axial coupling is of order $\Nc$ rather
than of order one.  Since the derivative of the pion field 
couples to the axial-vector
current, and in the potential, there are two such vertices, 
one can get a
long distance force of order $\Nc$.  Therefore the strong force due to pion
exchange at long distances and the large value of the axial coupling in
the Skyrme model are related.  

It is useful to understand the nature of the potential in the sigma model, due
to higher order pion exchanges.   First, let us look at the 
contributions to the
potential.  Note that if the vertices were not derivatively coupled, each pion
exchange would bring in a factor of $1/r$.  Due to the derivative coupling at
the vertices, there are two derivatives for each exchange.  There is also a
factor of $1/\fpi^2$.  This means the potential predicted by the non-linear
sigma model is of the form
\begin{equation}
  V_{\sigma} = {1 \over r} V_{\sigma}(\fpi r) .
\end{equation}
Note that this potential has the scale $R \sim 1/\fpi \sim
1/(\sqrt{\Nc}\lqcd)$,
so that it is much smaller than that of the standard nucleonic Skyrmion.

Perhaps it is easier to think about the pion field.  In the linear 
limit when we
treat the nucleon field as a point source, the pion field satisfies 
the equation
\begin{equation}
  -\nabla^2 \pi^a 
= {1 \over \fpi} \nabla^i \delta^{(3)} (\vec{r}) \la \sigma^i \tau^a \ra ,
\end{equation}
where $\sigma^i$ is a Pauli matrix.  This means that in lowest order, the pion
field is of the order of
$\pi \sim \fpi (1/r \fpi)^2$.  Higher corrections give
\begin{equation}
     \pi_\sigma = \fpi G_\sigma(\fpi r) .
\end{equation}
This is to be compared to that for the pion field of the Skyrme model
\begin{equation}
    \pi_{ {\rm Skyrme} } = \fpi G_{ {\rm Skyrme} }(\lqcd r) .
\end{equation}
We see that in the Skyrme case, that $\pi/\fpi \sim 1$ for $r \sim
1/\lqcd$ while for the sigma model this occurs at the much smaller
distance scale $r \sim 1/\fpi$.  

The axial vector coupling $g_A$ is estimated from the pion behavior 
at long distance,  $g_A\sim \fpi^2R^2$, where $R$ is the size of the pion could~\cite{Adkins:1983ya}.
In the Skyrme case, $g_A\sim \fpi^2/\lqcd^2\sim \Nc$, while in the sigma model
$g_A\sim \fpi^2/\fpi^2\sim 1$.

There are several subtleties in extracting the result for the Skyrmion case. 
Note that for any solution with a size scale $R \ll 1/\lqcd$, the
argument which led to the Skyrme term has broken down.
For such solutions, the Skyrme term itself is very small compared to the zeroth
order non-linear sigma model contribution at the size scale $1/\lqcd$. 
This is because in addition to the 
derivatives, there are four powers of the field, which are very small. 
Nevertheless, there should be a breakdown of the Skyrmion 
model at such a scale,
arising from QCD corrections of the underlying theory.  
The sigma model solution
sits at a distance scale small compared to where the Skyrmion action is
applicable, and one should ask what is the nature of the corrections to the
Skyrme model action at such distance scales.  In addition, there is good room
for skepticism about the Skyrme model treatment of the nucleon.   In the Skyrme
model, $1/e^2 \sim \Nc$, but phenomenologically is is of the order 
$3 \times 10^{-2}$.  If we were to naively take parametrically $1/e^2 \sim 1$,
the nucleon-nucleon force of the Skyrme model would be parametrically the same
as that in the sigma model.  The mass would not be correct however as it would
be of order $\fpi$.    In later sections we will see that this  picture has
features of what  we have in mind: a string vertex whose size is $1/\fpi$,
surrounded by a cloud of quarks.  

\section{What Might Be Wrong with the Skyrme Term?}

What might be wrong with the Skyrme model solution other than it it is not
consistent with
the sigma model?  Is there any problem with  internal inconsistency?
When attempts were made to derive the Skyrme term from QCD, one found
a Skyrme term which was generated, but also other terms
\cite{mackenzie,Aitchison:1986aq,Aitchison:1985yv,Aitchison:1984ys}.
When these terms were added together
and only terms of  fourth order in derivatives were retained,
the Skyrmion was found to be unstable to collapse.  
If this tendency to collapse
were maintained to all orders, then the Skyrmion might collapse to sizes much
less than the QCD distance scale.
One could not describe the nucleon within the conventional assumptions of the 
Skyrme model. 
(Strictly speaking, the Skyrme model comes from derivative 
expansion and keeping
only lowest order
terms is justified only when size scales 
$R \gg 1/\lqcd$ are considered.)

One can ask whether or not higher order terms might stabilize the Skyrmion. 
Following Refs. 
\cite{mackenzie,Aitchison:1986aq,Aitchison:1985yv,Aitchison:1984ys}, we
postulate that such corrections are generated by a quark determinant in the
presence of a background pion field.  We might hope that such a description
would be valid down to a scale of Skyrmion size of order $1/\fpi$.  It is at
this scale that high order terms in the pion nucleon sigma model generate
quantum corrections which are large.  This is also the natural size scale for a
pion since pion-pion interactions are of order $1/\Nc$, and even deep inelastic
scattering off of a pion is suppressed by $1/\Nc$.  Interactions with other
mesons are suppressed by $1/\Nc$.  If we assume that the quark-pion interaction
is parameterized by a vertex that is pointlike to a distance scale of order
$1/\fpi$, then this interaction strength is of order 
$g_{\pi QQ} \sim 1/\sqrt{\Nc}$, 
then one finds a contribution to the Skyrme term that is leading
order in $\Nc$.  This is because there are $\Nc$ quark loops.
Evaluating the leading term
in the derivative expansion of the pion field, 
there is the Skyrme term plus
two others, that have signs that cause the Skyrmion to collapse.  (It should be
noted that the intrinsic size scale over which quarks are 
distributed inside the
meson is more likely $1/\lqcd$, and the small apparent size of the 
pion arises from
the nature of interactions of these quarks in the large $\Nc$ limit, 
rather than
their intrinsic scale of spatial distribution.)

Subsequent to this \cite{Ripka:1985am}, it was argued that in chiral soliton
models, that
the chiral soliton is stable against collapse when the full quark
determinant is computed.  This happens when 
there are bound fermions in the presence of a nontrivial
background field,  and the energy of the bound quarks is
included.  This suggests that the Skyrmion could be
metastable if all orders in the fermion determinant are included.

We shall argue below that the Skyrmion at a size scale $R \sim 1/\lqcd$
is absolutely unstable.  Sufficiently small Skyrmions, $R \ll 1/\lqcd$
always collapse and they have an energy parametrically small 
compared to that of the nucleon.  

The quark contribution to the non-linear sigma model action modifies the
non-linear sigma model by
\begin{equation}
\delta S = \Nc ~\ln \{\det (  - i \Slash{\partial} + M \overline{U})\} .
\end{equation}
Here the quark mass $M$ is the constituent quark mass.  We now use Weinberg's
trick to rewrite this
as a coupling a to an axial-vector background field that is a pure gauge
transform of vacuum:
\begin{equation}
   A^\mu_5 = { 1\over i} V^{1/2} \partial^\mu V^{-1/2} .
\end{equation}   
Here $V^{1/2}$ is a function of the pion field and is a unitary matrix.  The
determinant becomes
\begin{equation}
   \delta S 
= \Nc \ln( \det( {1 \over i} \gamma^\mu(\partial_\mu - \gamma_5 
A_{\mu5}) + M)) .
\end{equation}

In the limit there $M = 0$, the fermion determinant is gauge invariant.  This
means that all functions
of $A$ generated by the determinant are gauge invariant and they vanish when
evaluated on 
$A^\mu_5$ which is a gauge transformation of vacuum field.

Now for fields that are slowly varying, this determinant may be computed by the
method of Refs.
\cite{mackenzie,Aitchison:1986aq,Aitchison:1985yv,Aitchison:1984ys}.
This yields the result that in leading order the Skyrmion collapses.  We also
see that if the Skyrmion is parametrically small compared to the scale size of
$\lqcd$, we can ignore the mass term in the fermion propagator, and no
potential is generated to resist the collapse of the Skyrmion.  
Since a Skyrmion
with size much less than $1/\lqcd$ has an energy 
arising from the non-linear sigma model contribution to the action that is
parametrically small compared to $\Nc \lqcd$, the 
Skyrmion is absolutely unstable.

It should be noted that it would be very difficult to resist the collapse on
very general grounds.  The collapse is prevented by fields that are singular at
short distances.  It is very difficult to generate
such singular terms on scale sizes much less than $1/\lqcd$ since QCD
interactions
are typically spread out on a distance scale of order $1/\lqcd$. The
exception to this is pion self-interactions, which are presumably special
because the pion is a Goldstone boson.
The reason for a lack of a short distance singularity on scale sizes much less
than $1/\lqcd$
is that the nucleonic core is color singlet and interactions that would produce
a $1/r$ singularity would need to couple to a non-zero color charge.  The
evasion to this conclusion  arises from quark kinetic energies.
If the quarks were confined to a size scale which is very small 
would generate a $1/R$ term.  

\section{The String Vertex Model and a Collapsed Skyrmion}

If the Skyrmion is unstable against collapse, a reasonable conjecture for its
minimum size is when quantum corrections to the
non-linear sigma model for the pions is large, when $R \sim 1/\fpi$.  This
limiting size can be understood from the Skyrme model itself.  Recall that the
energy of a Skyrmion of size $R$ is
\begin{equation} 
  E = \int ~d^3x \left[ {\fpi^2 \over { 16}} \tr~ \nabla U \nabla U^\dagger
\right] \sim \fpi^2 R .
\end{equation}  
Here we have ignored the possibility of a Skyrme term, since for small Skyrmions,
we have argued there is no  such term.  The energy of each constituent of the
Skyrmion is $1/R$, so that the number of quanta in the Skyrmion is
\begin{equation}
  N = \fpi^2 R^2 .
\end{equation}  
For a Skyrmion of size $R \sim 1/\lqcd$ this is $N \sim \Nc$  For the
collapsed Skyrmion, where
$R \sim 1/\fpi$, $N \sim 1$.  This is the limit where the 
quantum nature of the Skyrmion cannot be ignored.
 
The obvious problem with the collapsed Skyrmion is that it has a size
parametrically small compared to 
$1/\lqcd$.  On such size scales,  surely quark degrees of freedom are
important.
Since quarks carry a conserved charge they cannot be collapsed to small sizes
without paying a price
in kinetic energy $E \sim 1/R$, and so to keep the baryon mass from growing
larger than $\Nc \lqcd$, the quarks cannot be compressed 
to smaller than
the QCD scale.  Therefore if there is some remnant of the 
collapsed Skyrmion it
must include quarks and quark-antiquark pairs at the QCD size scale.  It is in
these degrees of freedom that the energy of the nucleon must reside.
The collapsed Skyrmion can only have an energy of order $\fpi$ and so does not
contribute much to the energy. 
 
\begin{figure}[htbp]
\begin{center}
\includegraphics[width=0.25\textwidth] {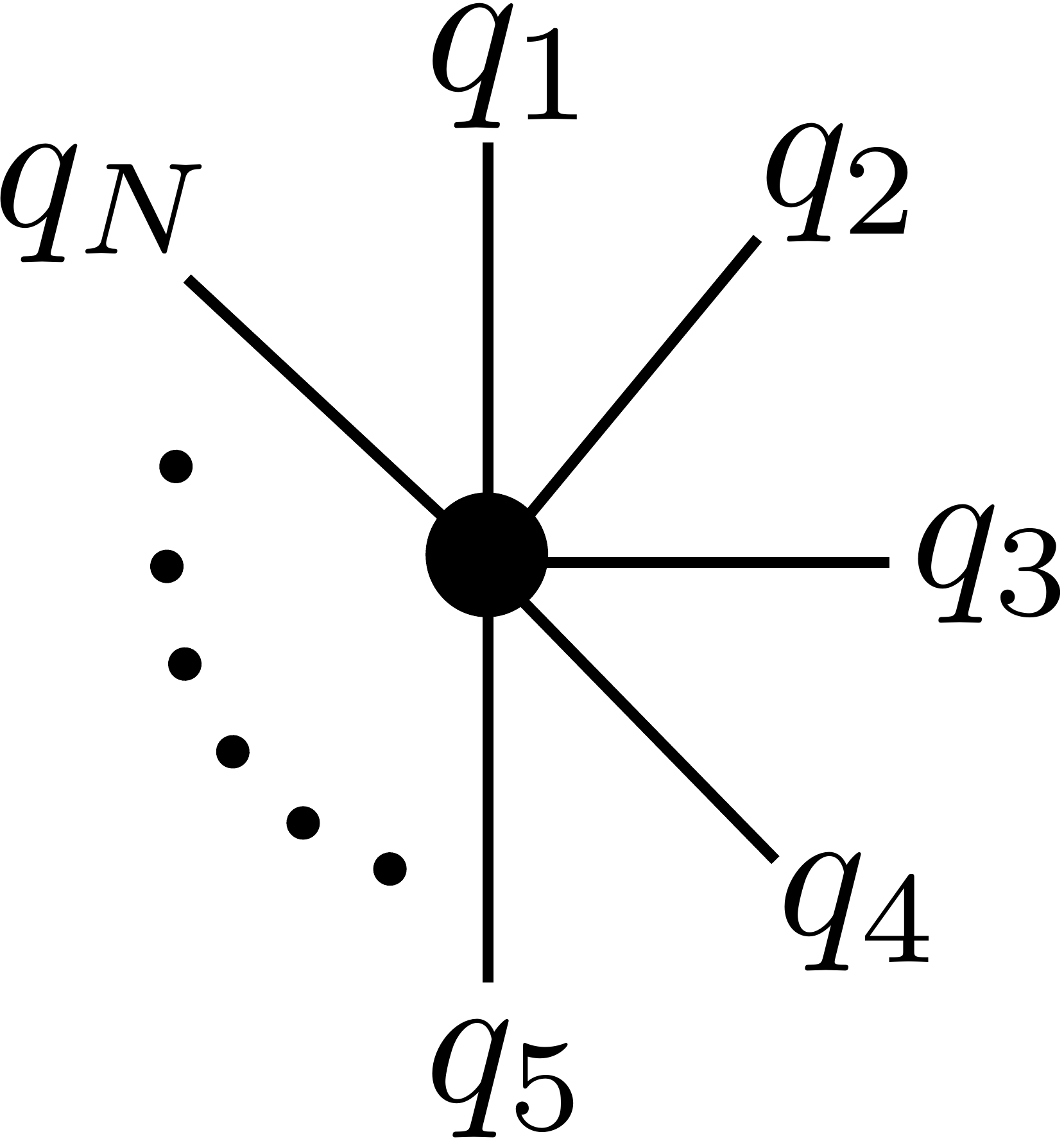}  
\end{center}
\caption{The topological baryon number operator of Veneziano.}
\label{veneziano}
\end{figure}

How can this picture of the nucleon be consistent with that of 
the quark model? 
Imagine that the nucleon is produced by the operator
\begin{equation}
      O_B (x) = \int \d^3x_1 \cdots d^3x_N q^{a_1}(x_1) U_{a_1,b_1}(x_1,x)
\cdots q^{a_N}(x_N) U_{a_N,b_N}(x_N,x) \epsilon_{b_1, \cdots , b_N} .
\end{equation}
Here, a path ordered phase along some path that connects the quark operator and
the position
of the baryon is denoted by $U(x,y)$.  This operator is the topological baryon
number operator of Veneziano \cite{Rossi:1977cy}.  It is shown pictorially in
Fig. \ref{veneziano}.

In this picture, quarks are joined together by lines of colored flux tubes at a
central point.  The quark operators are at a distance of order 
$1/\lqcd$
away from the central point.  We can identify the central point as the place
where the baryon number sits.  This is natural if we think about hadronizing
mesons along the lines of color flux.  This happens by 
quark-antiquark pairs and
so it is ambiguous to think about the baryon number as either centered at the
multiple string junction or at the ends of the strings.    As far as baryon
number is concerned, there is a symmetry between thinking about the baryon as
made of quarks or as a topological object is a fundamental dualism of the
theory:  We can either think about the baryon number as being delocalized on
quark degrees of freedom.  This is reflected in Cheshire Cat models of the
baryon \cite{Nadkarni:1984eg,Nadkarni:1985dn}.

In fact, it is easy to see that the degree of localization of the strong vertex
is the same as that of the collapsed Skyrmion.  Let us identify the string
vertex position with the average center of mass coordinate of the quarks,
\begin{equation}
  \vec{R} = {1 \over \Nc} (\vec{r}_1 + \cdots + \vec{r}_{\Nc}) .
\end{equation}
We work in a frame where $\la \vec{r}_i \ra = 0$.  The typical dispersion 
in the position of
the center of the string is therefore 
$\la \vec{R}^2 \ra \sim \la {\vec{r}_1}^{\;2} \ra /\Nc = 1/\fpi^2$. In
the Skyrmion picture, one imagines the collapsed Skyrmion as corresponding to
the string junction and having a high average density of baryon number
in a localized region, a picture that is dual to the quark model description.

The topological string model generates lines of colored electric flux from the
position of the string vertex.  This presumably results in linear 
confinement of
the quarks at distances far from the vertex.
Close to the vertex, each quark feels a strong color Coulombic 
interactions that
can be computed as the mean field of the color Coulombic fields of the other
quarks.  The Coulombic energy of all the quarks would be of order $\Nc/R$ times
the 't Hooft coupling, and the kinetic energy for relativistic quarks
would be of order $\Nc/R$.   The quarks sit at $R \sim 1/\lqcd$ 
in order not to make the nucleon energy larger than $\Nc \lqcd$.

\section{The Quark Distributions}

The picture above does not directly resolve the problems 
associated with a large axial-vector coupling or large matrix 
elements of the vector isospin currents.  
To understand what happens, we will take the non-relativistic quark 
model as a starting point.  
We will consider a matrix element of 
the non-relativistic expression of
axial-vector current, $\bar{q} \gamma_5 \gamma_3 \tau_3 q$,
which takes the form
\begin{eqnarray}
\la N| R_3 |N \ra \equiv 
 \la N| \sum_{q=1}^{\Nc}
    I_3^{(q)} S_3^{(q)} |N\ra ,
\end{eqnarray}
where the operator $O^{(q)}$ acts on $q$-th 
quark wavefunction contained in nucleon wavefunctions.

In nonrelativistic limit,
spins can characterize the irreducible representation
of Hamiltonian,
so wavefunctions can be characterized as
$|{\rm color}\ra \otimes |{\rm flavor}\ra \otimes |{\rm spin}\ra 
\otimes |{\rm space}\ra$.
Since color is totally antisymmetric,
we should totally symmetrize spin-flavor-space wavefunction.

The frequently used construction of baryon wavefunctions
is to use spin and isospin singlet diquark 
wavefunctions \cite{Kokkedee_book}.
We will denote a number of diquark pair as $n_d$.
We take a direct product of such a diquark state 
(and an extra quark if $\Nc$ odd),
then totally symmetrize spin-flavor-space wavefunctions.
In this construction, $\Nc=2n_d$ baryon is a spin-isospin singlet, 
while spin-isospin quanta of $\Nc=2n_d+1$ baryon
is solely determined by an extra quark.
There is nothing nontrivial when we 
compute matrix elements related to spin and isospin operators.

The situation differs for the computation of $\la R_3 \ra$.
The reason is that both of diquark and nucleon states
are not eigenstates of $R_3$, in contrast to
spin and isospin.
Below we will see this explicitly in terms of
$SU(2\Nf)$ representation of states.

To compute $\la R_3 \ra$, 
it is useful to use representations of 
the nonrelativistic $SU(4)$ symmetry \cite{Georgi_book}.
The $SU(4)$ algebra is formed by the following fifteen generators
\begin{eqnarray}
T_a = \sum_{q} I_3^{(q)} ,
\ \  
S_j = \sum_{q} S_j^{(q)} ,
\ \ 
R_{aj} = \sum_{q} T_a^{(q)} S^{(q)}_j ,
\end{eqnarray}
where $j=1,2,3$ and $a=1,2,3$.
The Cartan subalgebra of $SU(4)$
is formed by three generators
$I_3, S_3$ and $R_3\equiv R_{33}$, and 
states are characterized by
eigenvalues of these generators and 
the dimension $D$ of the irreducible representations.
We will denote such a state as
$|I_3,S_3,R_3;D\ra$.
In the following we often omit $D$ as far as
it brings no confusions.

As a preparation, 
let us express our spin-isospin singlet diquark state
in terms of $SU(4)$ representations.
It can be expressed as
\begin{eqnarray}
|D\ra 
&=& \frac{1}{2} \big\{ |u \up d \down \ra 
+ |d \down u \up \ra 
- |u \up d \down \ra 
- |d \down u \up \ra \big\}
 \nonumber \\
&=&
|0,0,1/2 \ra_{{\rm TS}} - |0,0,-1/2 \ra_{{\rm TS}},
\end{eqnarray}
where the subscript ${\rm TS}$ means that
wavefunctions are totally symmetrized.
Note that the diquark wavefuntion has
an eigenvalue of definite 3-component of
isospin and spin, while it is a mixture of
different $R_3$ eigenstates.
This is the origin to make
computations of $\la N|R_3|N\ra$ nontrivial.

Now we first argue simpler case, $\Nc =2n_d$ nucleons.
Such nucleons are characterized by states
with isospin and spin zero,
while they are mixture of states with different 
$R_3$ eigenvalues.
We assume that spatial wavefunctions are common for
all quarks, so that
spin-flavor (SF) wavefunctions of our nucleons
are obtained by 
totally symmetrizing a direct product of 
diquark's spin-flavor wavefunctions: 
\begin{eqnarray}
|N\ra_{2n_d}^{{\rm SF}}
= \big\{ \big(|0,0,1/2 \ra - |0,0,-1/2 \ra \big)^{n_d} \big\}_{ {\rm TS} }.
\end{eqnarray}
If we use $SU(4)$ expressions and omit subscripts on 
isospin and spin components (since they are zero),
\begin{eqnarray}
|N\ra_{2n_d}^{{\rm SF}}
&=& 
\big|\frac{n_d}{2} \big\ra_{ {\rm TS} } 
-{}_{n_d}\mathrm{C}_1 \big|\frac{n_d}{2}-1 \big\ra_{ {\rm TS} } 
+ \cdots 
+ (-1)^{n_d} \big|\frac{-n_d}{2} \big\ra_{ {\rm TS} } \nonumber \\
&=& 
\sum_{m=0}^{[n_d/2]} (-1)^{m} 
{}_{n_d}\mathrm{C}_m 
\bigg\{ \big|\frac{n_d}{2}-m \big\ra_{ {\rm TS} }
   +(-1)^{n_d} \big| - \big(\frac{n_d}{2} -m \big) 
\big\ra_{ {\rm TS} } \bigg\} ,
\end{eqnarray}
where $[n_d/2]$ equals to 
$n_d/2$ ($n_d/2-1$) for $n_d$ 
even (odd) case.
Now it is easy to see
\begin{eqnarray}
R_3 |N\ra_{2n_d}^{{\rm SF}}
&=& 
\sum_{m=0}^{[n_d/2]} (-1)^{m} 
{}_{n_d}\mathrm{C}_m 
\times \big( \frac{n_d}{2}-m \big) \nonumber \\
&\times&
\bigg\{ \big|\frac{n_d}{2}-m \big\ra_{ {\rm TS} }
   +(-1)^{n_d+1} \big| - \big(\frac{n_d}{2} -m \big) 
\big\ra_{ {\rm TS} } \bigg\}.
\end{eqnarray}
Note that relative sign in the second term 
is flipped after $R_3$ operation.
This gives $\la N | R_3 |N\ra_{2n_d}^{{\rm SF}}=0$
due to cancellations for each indices $m$.

From the above discussion, we saw that 
the matrix element of $R_3$ vanishes
not because nucleon states have small $R_3$,
but because cancellations occur
among contributions from different eigenstates.
Such cancellations are subtle, and strongly depend
on the fact that $\Nc$ is even.
Once we consider $\Nc$ odd baryons 
by adding one extra quark with the same spatial
wavefunction as others,
this situation completely changes:
terms which avoid cancellations lead to a huge value,
$\la N | R_3 |N\ra_{2n_d+1}^{{\rm SF}}=(\Nc+2)/12$.
So we now come back to the original problem,
a large value of $g_A$.

However, a comparison of $\Nc$ odd and even
baryons suggests us the following way 
to avoid a large $g_A$.
In the above, we always assume
all quarks occupy the same spatial wavefunction.
Here let us see what happens when we adopt 
a spatial wavefunction for the unpaired quark which is
different from that of quarks paired into diquarks.

We call spatial wavefunction of quarks inside
of diquark as $A(\vec{r})$, and
that of an extra quark as $B(\vec{r})$.
If we introduce a quantity $x$ which 
characterizes the overlap between 
wavefunction $A$ and $B$,
\begin{eqnarray}
x \equiv |\la A| B \ra| 
 = \big|\int d\vec{r} A^*(\vec{r}) B(\vec{r}) \big|,
\end{eqnarray}
then the expectation value of $R_3$ in the
$|p\up\ra$ state is
\begin{eqnarray}
\hspace{-0.5cm}
\la p\up |R_3|p\up \ra^{{\rm SFS}}
=
\frac{1}{12} \frac{ (\Nc-1)(\Nc+6)x^2 +12 }
  { (\Nc-1)x^2 + 4}.
\end{eqnarray}
The derivation is a little bit cumbersome,
so we give it in Appendix A.

Let us see the physical implications
of this result.
First the reason why $x^2$, not $x$,
appears is that a permutation of 
$A$ and $B$ always makes two $\la A|B \ra$.
For instance, $\la AAB|ABA \ra = x^2$.

For $x=1$, the matrix element reproduces 
conventional result, $(\Nc+2)/12$, as it should.

On the other hand, 
for $x=0$, or when $A$ and $B$ are
completely orthogonal each other,
the cancellations analogous to $\Nc=2n_d$ baryon
take place among the diquark part, so that $\la R_3 \ra$ is
merely characterized by $R_3$ for the leftover quark, $1/4$.

To get $g_A$ of $\sim \Nc^0$,
$x$ must be of order of $1/\Nc$.   Note also that $g_A$ is of order $\Nc$
until the overlap $x \sim 1/\sqrt{\Nc}$.  This 
means that in order to reduce $g_A$ from $\sim \Nc$, the overlap must be
very small.  This disparity in wavefunctions suggests the term
``dichotomous'' baryon.

\section{Why Small Overlap?}

Apparently the above small $g_A$ baryon
are not energetically favored in the 
shell model picture of quarks.
But so far we did not take into account
the contributions from fields surrounding 
valence quarks.
They affects the nucleon self-energy
via virtual mesonic loops
(polarization effects of the media).
If the axial charge of the valence quark 
is $\sim\Nc$, such a large charge induces 
a big change in the effective mass.

As argued above, we expect that the mass of the baryon is 
affected by a large value of $\Nc$.
A leading contribution to the $g_A$ dependent self-energy
comes from the vertex 
$\partial^\mu \pi_a/\fpi \times g_A \bar{N} \gamma_\mu \gamma_5 \tau_a N$,
and it generates $g_A^2/\fpi^2 \sim \Nc$ for 
$g_A \sim \Nc$~\cite{Dashen:1993as}
(no additional $\Nc$ dependence arises from nucleon propagator.)
We would expect other vector mesons might generate similar self-energies 
through coupling to the axial-vector current.  
From this self-energy dependence and 
the $x^2$ dependence of $g_A$, 
we suggests that self-energy effects generate a term like
\begin{equation}
 H_{g_A} \sim  \Nc ~  \mid \psi_{\rm paired}\mid^2 ~ \mid  \psi_{\rm unpaired} \mid^2,
\end{equation} 
in the effective Hamiltonian for the valence quarks.
Let us minimize this effective term.
Since there are of order $\Nc$ paired quarks, 
deforming their wavefunctions costs a lot of energy.
However, deforming the wavefunction of the unpaired quark only 
costs an energy of order $\lqcd$,
while the gain in $H_{g_A}$ is $\sim \Nc$.

This deformation is most easily accomplished by having 
the unpaired quarks exist in the region outside
of the paired quarks.  
The paired quark wavefunction in a string model falls
as $\exp{(-\kappa (r \lqcd)^{3/2})}$ at large distances.  
If the unpaired quark is excluded, 
due to its hard core interaction 
with the paired quarks, 
from a size scale $ r \le \ln^{2/3}(\Nc) /\lqcd$,
then $g_A$ can be reduced.

How large is the reduction in $g_A$?
If the self-energy is of order $g_A^2/\Nc$, 
then we would expect that when $g_A \sim \sqrt{\Nc}$, 
the trade off in energy associated with deforming the unpaired quark 
wavefunction is balanced by self-energy effects.  
Such a reduction most likely allows for a phenomenologically
acceptable large $\Nc$ limit.  

Magnetic moments have been computed in Appendix B,
and are proportional to $g_A$. 
Magnetic interactions will be of order $\alpha_{{\rm em}} g_A^2$, 
so that for sufficiently large $\Nc \sim 1/\alpha_{{\rm em}}$, these
effects would also work toward reducing $g_A$ to a value of order $1$.

It is also possible that a large value of $g_A$ might mean even more singular
self-energy terms for large $\Nc$ resulting 
is a greater reduction of $g_A$.  
For example, there might 
in principle be effects that contribute to the
energy that  correspond to higher powers than linear in $\Nc$ when 
$g_A \sim \Nc$.  These terms will tend to further 
reduce the parametric dependence of $g_A$ upon $\Nc$. 
We have not been able to find a compelling argument for such effects 
from strong interactions, though.

We turn next to a discussion of the splitting between the nucleon and
the $\Delta$.
In a conventional non-relativistic quark model, there is a 
$SU(2 \Nf)$ symmetry which requires the $N-\Delta$ masses to be equal
to $\sim \lqcd$.  This degeneracy is 
split by color hyperfine interaction,
\begin{eqnarray}
\sum_{i\neq j} V_{{\rm ss}}(\vec{r}_{ij}) 
\sim 
\frac{\lambda}{\Nc} \sum_{i\neq j}
\frac{\vec{S}_i \cdot \vec{S}_j}{M_i M_j} \delta(\vec{r_{ij}});
\end{eqnarray}
$\lambda = g^2 \Nc$, and the $M_i \sim 1$ are constituent quark masses.
Assuming all quark masses are the same,
the expectation value for a state with spin $S$ is
\begin{eqnarray}
\la \sum_{i\neq j} V_{{\rm ss}}(\vec{r}_{ij}) \ra
\sim 
\frac{\lambda}{\Nc} 
\big[S(S+1) - \frac{3}{4} \Nc\big] 
\times 
\frac{|\phi_{\rm relative}(\vec{0})|^2}{M^2}.
\end{eqnarray}
Masses are split by the first term.  If the difference in spins
is $\sim 1$, as for the nucleon and 
the $\Delta$, the mass splitting is $\sim 1/\Nc$.
This agrees with the Skyrme model, identifying the
$\Delta$ as the first spin excitation of the nucleon.
More general arguments can be found in \cite{Dashen:1993as,Carone:1993dz}.

In contrast, there is no $SU(2\Nf)$ symmetry in the model of
a Dichotomous Baryon.  The masses of the nucleon and the $\Delta$
are not nearly degenerate, but split $\sim \lqcd$.
This arises from polarization effects via the axial coupling of the $\Delta$,
$g_{\Delta A}$.

Consider what a dichotomous $\Delta$, with $I_3=S_3=3/2$, is like.
This can be obtained by breaking apart one diquark pair:
\begin{eqnarray}
|D\ra \longrightarrow
|D'\ra = \frac{1}{\sqrt{2} } 
\big\{ |u \up u \up \ra + |u \up u \up \ra  \big\}
=|1,1,1/2 \ra_{ {\rm TS} }.
\end{eqnarray}
Suppose that these $|u\up \ra$ occupy the same 
spatial wavefunctions as those in the diquark pairs.
Then $g_{\Delta A}$ is $\sim\Nc$, and $\Delta$ has
a large vacuum polarization of $\sim\Nc$.  As with the unpaired quark
in the nucleon, this can be avoided by putting both $u$ quarks into
a spatial wavefunction which is orthogonal to that of the paired quarks.
This costs an excitation energy $\sim \lqcd$,
not $\sim \lqcd/\Nc$.

If $M_\Delta - M_N\sim \lqcd$ and  
$g_{\Delta N A}\sim1$, the width of the $\Delta$ is
\begin{equation}
\Gamma_\Delta\sim \frac{g_{\Delta N A}^2}{\fpi^2}
\left(\frac{M_\Delta^2-M_N^2}{M_\Delta}\right)^3
\sim \frac{g_{\Delta N A}^2}{\Nc} \, \lqcd\; .
\end{equation}%
Thus whether $g_{\Delta N A}$ is $\sim 1$ or $\sim \sqrt{\Nc}$,
the $\Delta$ remains a narrow resonance at large $\Nc$.
We note that in QCD, the $\Delta$ is not broad,
$\Gamma_\Delta\sim 118 $MeV~\cite{PDG}.
By the Adler-Weisberger relation~\cite{adler},
$g_A$ is of the same order as $g_{\Delta N A}$ \cite{Mazzitelli:1986hg}.

\section{Summary, and tests on the lattice}

In this paper we considered the properties of
baryons for a large number of colors.
It is certainly possible that our analysis only applies for an
unphysically large value of $\Nc$.  

The question of how $g_A$ grows with $\Nc$ has important implications
for nuclear physics, though.  The central conundrum of nuclear physics
is that the binding energy of nuclear matter is much smaller than any
other mass scale in QCD.  This is usually explained as the result of
a nearly exact cancellation between repulsion, from the exchange of
$\omega$-mesons, and attraction, from $\sigma$-exchange.

In QCD, the $\sigma$ meson is light, $\sim 600$~MeV, and very broad.
This may be because for three colors, the $\sigma$ meson is really
a state involving two quark anti-quark pairs \cite{alford_jaffe}.  

For a large number of colors, though, the lightest scalar meson only
has a single quark anti-quark pair, and is probably heavy, 
with a mass significantly greater than that of 
the $\omega$-meson \cite{largeN_sigma}.  In this case, there cannot
be any approximate cancellation between $\omega$
and $\sigma$ exchange: for distances greater than the inverse mass of the
$\omega$, there is just repulsion, $\sim \Nc$.  

For distances greater than $\sim \log(\Nc)/m_\omega$, then, the only
interaction is due to pions.  This can be analyzed in chiral perturbation
theory \cite{weise}, where two pion exchange gives a result $\sim g_A^2$.
If $g_A \sim \Nc$, this looks too large, but because of the cancellations
of Dashen and Manohar \cite{Dashen:1993as}, the nucleon-nucleon potential
is only $\sim \Nc$.  

If, however, the present analysis is correct, and $g_A$ does not grow with
$\Nc$, then the nucleon-nucleon potential is much smaller, $\sim 1/\Nc$.
Thus having a value of the axial coupling which does not grow with $\Nc$
may help understand why nuclear matter is so weakly bound.
The price we have to pay is that
we may have to give up 
the elegant contracted spin-flavor 
$SU(2\Nf)$ symmetry
which was derived under the assumption
of $g_A \sim \Nc$ \cite{Dashen:1993as}.

One glaring shortcome of our analysis is 
that we only consider two light flavors.  
An extension of our diquark based construction
to three flavor cases is not straightforward.
While diquarks behave as singlets in two flavors,
they are anti-triplet in
the $SU(3)$ flavor representations.
It means that diquarks have some charges
under the $SU(3)$ flavor symmetry,
which should be cancelled to reduce Goldstone boson clouds.
Thus we have to carefully combine diquarks, or instead
it may be better to look for other basic ingredients
which play similar roles as diquarks in 
two flavor theories.

In this paper, we implicitly took a view that
the strange quark is heavy enough to
suppress kaon clouds, so 
we did not try to reduce the nucleon-kaon axial couplings.
An obvious problem is that this treatment
badly breaks the $SU(3)$ flavor symmetry,
which explain
mass splittings among octet or decouplet baryons
by regarding the strange quark mass as a perturbation.  
To reduce these gaps,
we plan to study the $SU(3)$ chiral limit, 
which will be discussed elsewhere.

On the other hand, the questions which we raise for two flavors
can be addressed, at least indirectly, by
numerical simulations on the lattice.  We thus conclude by discussing
these results, and suggest further study.  

The spectrum of baryons has been studied on the lattice 
\cite{lattice_baryon_spectrum}.  While the simulations are for
quark masses with pions which are significantly heavier than the
physical pion mass, there are striking differences from the observed
spectrum of baryons in QCD.  In particular, there are several states
which are present in QCD, but not on the lattice with heavy quarks.
Notably, this includes the Roper resonance $N(1440)$, as well as
other states.  This is a puzzle for a
non-relativistic, constituent quark model.  

In this paper we have not considered baryon excitations, and so have
not addressed the problem of the Roper resonance, or other
similar states.  We find it intriguing,
however, that at present results from the lattice appear to demonstrate
that some states in the baryon spectrum are very sensitive to the
chiral limit.  

Lattice simulations have also measured the axial charge of the baryon
\cite{lattice_axial_charge}.  These results show that even for very heavy
pion masses, $m_\pi \sim .7$~GeV, that the axial charge of the nucleon is 
much smaller than the value of the constituent quark model, 
$g_A \sim 1.2$.  For lighter pion masses, the axial charge decreases to 
a value near one.  Again, such a sensitivity of the axial charge to
the chiral limit is unexpected in a non-relativistic, constituent
quark model.

Besides simulations with three colors, it would also be of interest to
simulate baryons for five colors  \cite{ohta}.  
(It is necessary to take the number of
colors to be odd, so that the lightest nucleon has nonzero spin.)  
Even in the quenched approximation, it would be interesting to know
if the axial charge is close to the value in the non-relativistic 
quark model, $= 7/3$, or to unity.

\section{Acknowledgments}

L. McLerran thanks Tom Cohen and Dmitri Diakonov for heated discussions on this
subject, and Ismail Zahed for critical observations.  He 
gives enormous thanks to 
Jean-Paul Blaizot and Maciej Nowak, with whom he had many 
discussions in the early stages of this project; he also thanks
hospitality of the Theoretical Physics Division of CEA-Saclay, where 
this work was initiated.  We also thank Y. Aoki, K. Hashimoto,
D. K. Hong, D. Kaplan, M. Karliner, K. Kubodera,
S. Ohta, M. Rho, S. Sasaki, and M. Savage 
for discussions and comments.
T. Kojo is supported by Special Posdoctoral Research Program of RIKEN;
he also thanks the Asia Pacific Center for
Theoretical Physics for their hospitality during a visit in
June, 2010.
This manuscript has been authorized under Contract No. DE-AC02-98CH0886 
with the US Department of Energy.
This research of Y. Hidaka is supported by the Grant-in-Aid for
the Global COE Program ``The Next Generation of Physics, 
Spun from Universality and Emergence'' from the Ministry of 
Education, Culture, Sports, Science and Technology (MEXT) of Japan.

\appendix
\section{Computation of $g_A$}

The purpose of this appendix is to reproduce
a well-known results of 
$\la R_3 \ra = 1/4 \times (\Nc+2)/3$ for
$\Nc$ odd nucleons which is composed of 
nonrelativistic quarks occupying the same space wavefunction.
We will also extend this result to
the case with different space wavefunctions.

If we assume that all quarks have the same
space wavefuntion, we have only to completely symmetrize
spin-flavor wavefunctions to satisfy the Pauli's principle.
Here we will consider $|p\up \ra$
which, in our consturction, takes the form:
\begin{eqnarray}
|p\uparrow \ra^{{\rm SF}}
= \big\{ \big(|0,0,1/2 \ra - |0,0,-1/2 \ra \big)^{n_d}
 \otimes |u\up \ra
\big\}_{ {\rm TS} }.
\end{eqnarray}
As in the text,
we will omit third component of spin and isospin
of the diquark wavefunction for notational simplicity.
The expression is
\begin{eqnarray}
|p\up \ra^{{\rm SF}}
= \sum_{m=0}^{n_d} (-1)^{m} 
{}_{n_d}\mathrm{C}_m  
\bigg\{ \big| \frac{n_d}{2}-m \big\ra \otimes |u\up \ra
\bigg\}_{ {\rm TS} }.
\end{eqnarray}
First we have to give a correct normalization.
It is crucial to count a number of independent states 
which are contained in maximally symmetrized $R_3$ eigenstates.

For instance, 
$|u\up u\up u\up \ra_{ {\rm TS}}$
has only one independent state, and
degeneracy factor $3!$ for symmetrization,
so $ \la u\up u\up u\up |u \up u \up u \up \ra_{ {\rm TS}}
=(3!)^2 = 36$. 
On the other hand,
$|u\up u\up d\up \ra_{ {\rm TS}}$
has three independent states and degeneracy factor 
$3!/3 = 2!$, 
so 
$ \la u\up u\up d\up |u \up u \up d \up \ra_{ {\rm TS}}
= 3 \times (2!)^2 = 12$.

Since our diquark state $|0,0,1/2\ra$ contains 
$(u \up, d \down)$ and 
$|0,0,-1/2\ra$ contains $(u\down, d\up)$,
a state $\{ |n_d/2 -m \ra \otimes |u \up \ra \}_{{\rm TS}}$ 
includes
$(n_d-m+1)$ number of $u\up$,
$(n_d-m)$ number of $d\down$,
and $m$ number of $u\down$ and $d\up$.
This state can be written in the following way:
\begin{eqnarray}
\hspace{-0.5cm}
\bigg\{ \big| \frac{n_d}{2}-m \big\ra \otimes |u\up \ra
\bigg\}_{ {\rm TS} }
= D(n_d-m+1;n_d-m;m;m)
\big\{ |u \up, ...\ra + ...
\big\} ,
\end{eqnarray}
where 
$D(n_d-m+1;n_d-m;m;m) = (n_d-m+1)! (n_d-m)! (m!)^2$
is a degeneracy factor
coming from multiple counting of the same
quarks.
A number of independent states in the bracket, 
$C(n_d-m+1;n_d-m;m;m)$,
is given by 
a number of permutation $(2n_d+1)!$
devided by degeneracy factor $D$.
From these observations,
we get
\begin{eqnarray}
\hspace{-0.5cm}
\big\la \big( \frac{n_d}{2}-m \big)\otimes u\up 
\big| \big( \frac{n_d}{2}-m \big) 
 \otimes u\up \big\ra_{ {\rm TS} }
= D^2 \times C = (2n_d+1)! \times D.
\label{normalization}
\end{eqnarray}
Now normalization factor of $|p \up\ra^{{\rm SF}}$
can be computed as
\begin{eqnarray}
\la p\up |p\up \ra^{{\rm SF}}
&=& 
(2n_d+1)! \sum_{m=0}^{n_d}
( {}_{n_d}\mathrm{C}_m )^2 \times
D(n_d-m+1;n_d-m;m;m) \nonumber \\
&=&
(2n_d+1)! (n_d!)^2 \times \frac{(n_d+1)(n_d+2)}{2}.
\end{eqnarray}
Now computation of 
$\la p\up |R_3|p\up \ra^{{\rm SF}}$
is straightforward.
We have only to multiply an eigenvalue
$(n_d/2-m +1/4)$ when we take the sum of
$m$,
\begin{eqnarray}
\la p\up |R_3|p\up \ra^{{\rm SF}}
&=& 
(2n_d+1)! \sum_{m=0}^{n_d} 
\big(\frac{n_d}{2}-m +\frac{1}{4} \big)
\times
( {}_{n_d}\mathrm{C}_m )^2 \times D
\nonumber \\
&=&
(2n_d+1)! (n_d!)^2 \times \frac{(n_d+1)(n_d+2)(2n_d+3)}{24}.
\end{eqnarray}
Using $\Nc=2n_d+1$, we reproduce the well-known result, 
$\la p\up |R_3|p\up \ra^{{\rm SF}}/\la p\up |p\up \ra^{{\rm SF}}
=(\Nc+2)/12$.

Next we little bit extend the results 
to the case where all diquark wavefunctions
occupy the same space wavefunction, $A(\vec{r})$, while
extra quark occupies a different space
wavefunction, $B(\vec{r})$.
In such a case, it is no longer useful to
separate treatments of spin-flavor and space.
Rather we will totally symmetrize
spin-flavor-space (SFS) wavefunctions with
explicitly expressing space dependence in such a way that
$|u\up,A\ra $, $|u\up,B \ra$,.., and so on.

Here we introduce a quantity $x$
which characterizes the overlap between 
wavefunction $A$ and $B$,
\begin{eqnarray}
x \equiv |\la u\up,A| u\up,B \ra|,
\end{eqnarray}

We will consider the following nucleon states
\begin{eqnarray}
|p\up \ra^{{\rm SFS}}
= \sum_{m=0}^{n_d} (-1)^{m} 
{}_{n_d}\mathrm{C}_m  
\bigg\{ \big| \frac{n_d}{2}-m, A \big\ra \otimes |u\up,B \ra
\bigg\}_{ {\rm TS} }.
\end{eqnarray}
where
\begin{eqnarray}
\hspace{-0.5cm}
&& \bigg\{ \big| \frac{n_d}{2}-m,A \big\ra \otimes |u\up,B \ra
\bigg\}_{ {\rm TS} } \nonumber \\
&& = D(n_d-m;n_d-m;m;m) \times
\big\{ |u \up,A; \cdots ; u \up,B \ra + ...
\big\} .
\label{normalization2}
\end{eqnarray}
We distinguish $|u\up,A\ra$ and $|u\up,B\ra$,
so that, compared to previous case,
the degeneracy factor $D$ decreases 
by a factor $1/(n_d-m+1)$ while 
a number of independent states $C$ increases
by a factor $(n_d-m+1)$.

Now to see how nonzero overlap of $A$ and $B$ arises,
let us take the innner-product of braket in 
(\ref{normalization2}):
\begin{eqnarray}
&& \big\{ \la u \up,A; \cdots ; u \up,B| + ...
\big\}
\big\{ |u \up,A; \cdots ; u \up,B \ra + ...
\big\} \nonumber \\
&&
= C(n_d-m;n_d-m;m;m)
\times \big\{1 + x^2 \times (n_d-m) \big\}.
\end{eqnarray}
The first term comes from diagonal matrix elements,
while second term comes from offdiagonal terms.
(Perhaps the simplest way to determine
a coefficient of $x^2$ is to see 
that $x=1$ reproduce the previous results 
(\ref{normalization}) ).

Remaining calculations are just a repetition
of the previous calculations.
A normalization factor is
\begin{eqnarray}
\la p\up |p\up \ra^{{\rm SFS}}
&=& 
(2n_d+1)! \sum_{m=0}^{n_d}
( {}_{n_d}\mathrm{C}_m )^2 \times
D(n_d-m;n_d-m;m;m) \nonumber \\
&& \hspace{2cm}
   \times \big\{ 1+x^2 (n_d-m) \big\}
\nonumber \\
&=&
(2n_d+1)! (n_d!)^2 \times \frac{(n_d+1)(x^2 n_d+2)}{2},
\end{eqnarray}
which of course reproduces previous results for $x=1$.
And also the expectation value of $R_3$ is
\begin{eqnarray}
\hspace{-0.5cm}
\la p\up |R_3|p\up \ra^{{\rm SFS}}
&=& 
(2n_d+1)! \sum_{m=0}^{n_d}
( {}_{n_d}\mathrm{C}_m )^2 \times
D(n_d-m;n_d-m;m;m) \nonumber \\
&& \hspace{2cm}
   \times \big\{ 1+x^2 (n_d-m) \big\} 
   \times (n_d/2-m +1/4)
\nonumber \\
&=&
(2n_d+1)! (n_d!)^2 
 \times \frac{(n_d+1)(2n_d^2 x^2 +7n_d x^2 +6 )}{24},
\end{eqnarray}
Finally, taking into account the normalization factor,
we get
\begin{eqnarray}
\hspace{-0.5cm}
\frac{ \la p\up |R_3|p\up \ra^{{\rm SFS}} }
{ \la p\up |p\up \ra^{{\rm SFS}} }
=
\frac{2n_d^2 x^2 +7n_d x^2 +6}{ 12(x^2 n_d+2)}
=
\frac{1}{12} \frac{ (\Nc-1)(\Nc+6)x^2 +12 }
  { (\Nc-1)x^2 + 4}.
\end{eqnarray}
%
\section{Magnetic Moments}
%
The purpose of this section is 
to give a relationship between
$\la R_3 \ra$ and magnetic moments:
\begin{eqnarray}
\mu_N
&=& 
\la N | \big[ \sum_u \mu_u S_3^{(u)}
 + \sum_d \mu_d S_3^{(d)} \big] |N \ra .
\end{eqnarray}
We assume $|N\ra$ for spin $\up$ case
in the following.
By introducing isospin projector,
the sum of ($u,d$) indices can be extended, 
so we can rewrite the sum in terms of 
total spin and $R_3$ operators,
\begin{eqnarray}
\mu_N
&=& 
\la N | \big[ \sum_q \mu_u S_3^{(q)}
 \big( \frac{1}{2} + I_3^{(q)} \big) 
 + \sum_q \mu_d S_3^{(q)} 
    \big( \frac{1}{2} - I_3^{(q)} \big)
\big] |N \ra 
\nonumber \\
&=& 
\frac{\mu_u + \mu_d}{2} 
\la N | \sum_q S_3^{(q)} |N \ra 
 +
(\mu_u - \mu_d)
\la N | \sum_q R_3^{(q)} |N \ra 
\nonumber \\
&=& \frac{
(\mu_u + \mu_d) \pm g_A (\mu_u -\mu_d) }{4},
\end{eqnarray}
where $+$ ($-$) signs for protons (neutrons).

Assuming consistuent masses of ($u,d$) quarks
are almost same, and using
$Q=I_3+B/2=I_3+1/2\Nc$, we denote
quark magnetic moments $Q\bar{\mu}\equiv Qe/2M_q$:
\begin{eqnarray}
\mu_u = \frac{\Nc+1}{2\Nc} \bar{\mu}, \hspace{1cm}
\mu_d = -\frac{\Nc-1}{2\Nc} \bar{\mu}.
\end{eqnarray}
Therefore proton and neutron magnetic moments are
\begin{eqnarray}
\mu_{p,n} = \frac{\bar{\mu}}{4}
  \bigg( \frac{1}{\Nc} \pm g_A \bigg).
\end{eqnarray}
For conventional baryon wavefunctions,
$\Nc=3$ and $g_A=5/3$,
which gives $\mu_n/\mu_p= -2/3$ (exp:$-0.685$).
For our wavefunction with $x^2=0$,
$g_A=1$ and $\mu_n/\mu_p= -1/2\ (-1)$ for $\Nc=3\ (\infty)$.


\end{document}